\documentclass[aip,jcp,amsmath,amssymb,floatfix,reprint,citeautoscript]{revtex4-2}

\usepackage[utf8]{inputenc}
\usepackage{graphicx}
\usepackage{wrapfig}
\usepackage[colorlinks,allcolors=black,citecolor=blue,urlcolor=blue]{hyperref}
\usepackage[mathlines]{lineno}
\usepackage{ae}
\usepackage{amsmath}
\usepackage{booktabs}
\usepackage{silence}
\graphicspath{{figures/}}

\begin{document}

\def\mytitle{Benzene Radical Anion Microsolvated in Ammonia Clusters: Modelling the Transition from an Unbound Resonance to a Bound Species}
\title{\mytitle}

\author{Vojtech Kostal}
\affiliation{
Institute of Organic Chemistry and Biochemistry of the Czech Academy of Sciences, Flemingovo nám. 2, 166 10 Prague 6, Czech Republic
}

\author{Krystof Brezina}
\affiliation{
Charles University, Faculty of Mathematics and Physics, Ke Karlovu 3, 121 16 Prague 2, Czech Republic
}
\affiliation{
Institute of Organic Chemistry and Biochemistry of the Czech Academy of Sciences, Flemingovo nám. 2, 166 10 Prague 6, Czech Republic
}

\author{Ondrej Marsalek}
\affiliation{
Charles University, Faculty of Mathematics and Physics, Ke Karlovu 3, 121 16 Prague 2, Czech Republic
}

\author{Pavel Jungwirth*}
\email{pavel.jungwirth@uochb.cas.cz}
\affiliation{
Institute of Organic Chemistry and Biochemistry of the Czech Academy of Sciences, Flemingovo nám. 2, 166 10 Prague 6, Czech Republic
}

\date{\today}

\begin{abstract}

\setlength\intextsep{0pt}
\begin{wrapfigure}{r}{0.40\textwidth}
  \hspace{-1.8cm}
  \includegraphics[width=0.40\textwidth]{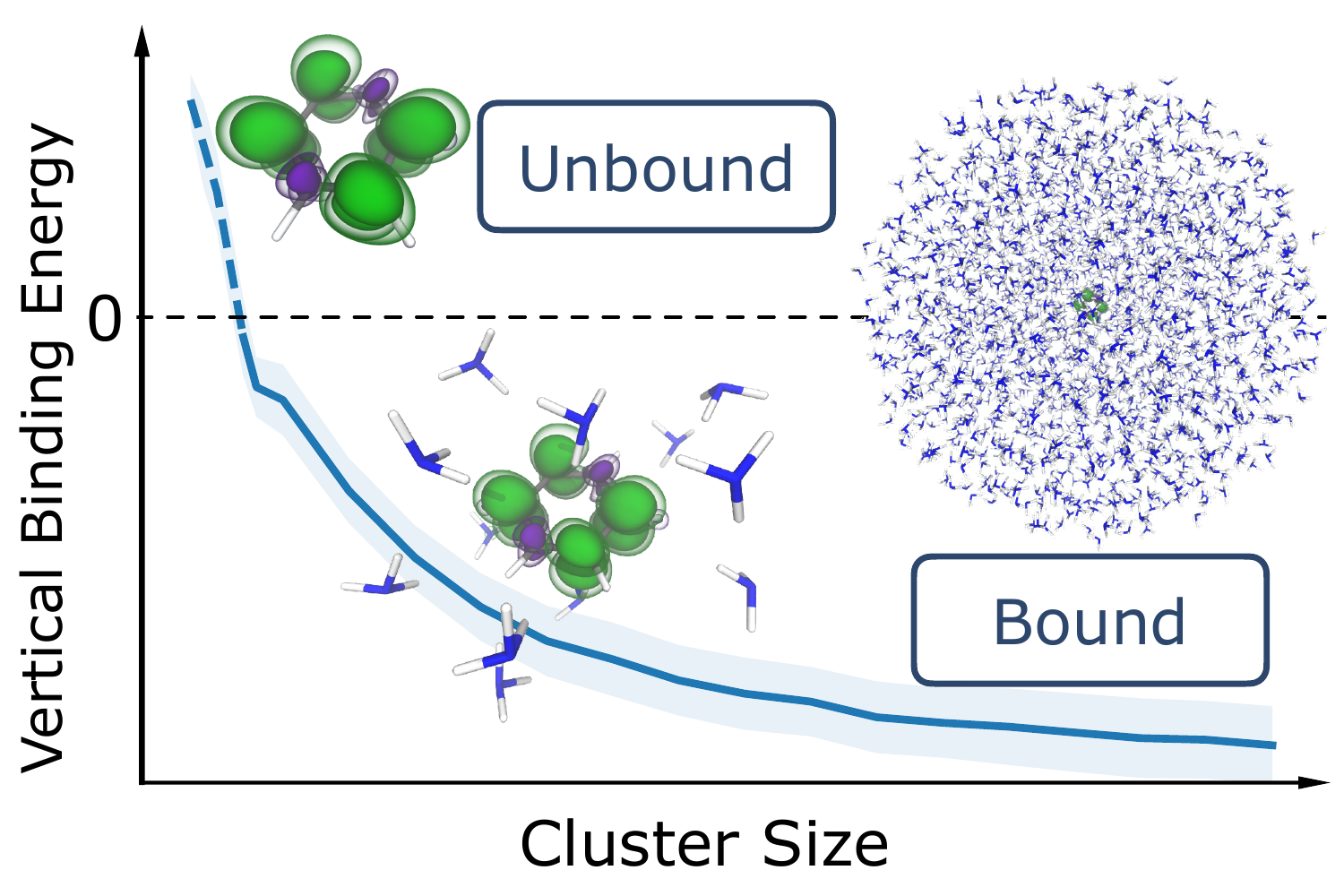}
\end{wrapfigure}

The benzene radical anion, well-known in organic chemistry as the first intermediate in the Birch reduction of benzene in liquid ammonia, exhibits intriguing properties from the point of view of quantum chemistry.
Notably, it has the character of a metastable shape resonance in the gas phase, while measurements in solution find it to be experimentally detectable and stable.
In this light, our previous calculations performed in bulk liquid ammonia explicitly reveal that solvation leads to stabilization.
Here, we focus on the transition of the benzene radical anion from an unstable gas-phase ion to a fully solvated bound species by explicit ionization calculations of the radical anion solvated in molecular clusters of increasing size.
The computational cost of the largest systems is mitigated by combining density functional theory with auxiliary methods including effective fragment potentials or approximating the bulk by polarizable continuum models.
Using this methodology, we obtain the cluster size dependence of the vertical binding energy of the benzene radical anion converging to the value of $-$2.3~eV at a modest computational cost.

\end{abstract}

\maketitle

\section{Introduction}

The Birch reduction~\cite{Birch1946}, typically realized in liquid ammonia in the presence of simple alcohols, is a synthetic method of reducing aromatic compounds that offers a high level of control and selectivity by employing alkali metals to convert benzene derivatives to 1,4-cyclohexadienes.
It gained significant importance in the area of organic synthesis owing to its applicability to reduce a wide range of aromatic substrates including highly stable molecules such as benzene~\cite{Hook1986}.
The reaction mechanism relies on the spontaneous generation of solvated electrons by dissolution of alkali metals in liquid ammonia~\cite{Zurek2009, Buttersack2020} that then bind to the substrate to create a radical anion species.
Specifically, in the case of benzene, which represents the simplest substrate that can undergo the Birch reduction, the benzene radical anion is formed.

In the gas phase, the benzene radical anion is not electronically stable and its life time was experimentally found to be on a femtosecond time scale~\cite{Sanchef1973}.
This was supported by coupled cluster calculations that revealed a positive value of the vertical binding energy (VBE) or, equivalently, a negative ionization potential reflecting the character of a metastable shape resonance~\cite{Bazante2015}.
In contrast, spectroscopic measurements performed in solution~\cite{Tuttle1958, Ishitani1967, Moore1981} imply that the benzene radical anion must be stable in the solvated environment to be experimentally detectable on extended time scales.
These experiments also show that the stability of the benzene radical anion is not conditioned specifically by liquid ammonia, but rather it is achieved in a multitude of polar solvents such as dimethoxyethane~\cite{Mortensen1984} or tetrahydrofuran~\cite{Marasas2003}.
Moreover, note that the feasibility of the Birch reduction itself also requires stability of the intermediate in order to participate in a chemical reaction.

A molecular-level insight into the stability of such systems is provided by mass spectrometry and photoelectron spectroscopy measurements of microsolvated clusters showing that benzene and naphthalene radical anion-water clusters are stabilized, in terms of VBE, already by a small number of water molecules~\cite{Maeyama1997, Lyapustina2000}.
In general, cluster experiments offer a flexible way to explore solvation effects on the boundary between the gas phase and the solution, as illustrated by the study of sodium atoms in ammonia vapor~\cite{Steinbach2005} or the sulfate dianion in water~\cite{Wang2009}.
In this work, we consider ammonia only, since it is the solvent of choice in the context of the Birch reduction due to its remarkable ability to sustain solvated electrons for extended periods of time~\cite{Zurek2009}, but note that there is a qualitative parallel between water and ammonia as a pair of related hydrogen-bonded liquids.

Theoretical calculations can provide meaningful insight into the solvent effects on the electronic stability of the benzene radical anion that complement the experimental point of view.
However, the open-shell and diffuse character of the radical anion requires a high level of electronic structure theory which, in combination with the extensive modeling of the explicit bulk solution, renders the calculations computationally demanding. 
In this direction, we recently reported a hybrid density function theory (DFT) \textit{ab initio} molecular dynamics (AIMD) simulation of the benzene radical anion in liquid ammonia in periodic boundary conditions at 223~K~\cite{Brezina2020}.
There, the benzene radical anion was found to retain a stable spin population over the course of the whole simulation which points to electronic stability in terms of localization of the excess electron on the aromatic ring.
In a follow-up study, the one-electron binding energies were calculated for the AIMD bulk geometries employing the accurate G$_0$W$_0$~\cite{Huser2013, Wilhelm2016} electronic structure method~\cite{Brezina2021} again in periodic boundary conditions.
These results explicitly show that the solvated radical anion is electronically stable and also provide a VBE value of $-$2.3~eV relative to the vacuum level.

The \textit{ab initio} studies aimed so far at two extreme situations---the electronically unbound isolated species~\cite{Bazante2015} on one hand, and the fully solvated stable system on the other hand~\cite{Brezina2020, Brezina2021}.
Here, we bridge the gap by explicit ionization calculations employing hybrid DFT performed on clusters of increasing size.
First, we address the bulk value of the VBE of clusters embedded in a polarizable continuum~\cite{Miertus1981, Tomasi2005} employing a polarizable continuum model (PCM) augmented by a small explicit solvent region to account for local solvent--solute interactions.
This approach provides a baseline at a relatively modest computational cost and was used successfully in previous computational investigations of both ionization~\cite{Cauet2010, Ghosh2011, Toth2020} and excitation~\cite{Provorse2016} phenomena in solution.
Next, we remove the PCM and study clusters of various sizes in the gas phase to describe the stabilization process that takes place when the isolated species is gradually surrounded by an increasing number of solvent molecules.
There are certain benefits to this approach---it provides both the smallest number of solvent molecules necessary to stabilize the solute as well as the extent of solvation needed to converge the VBE to their bulk values.
This convergence was shown to be typically very slow, in particular with charged solutes as illustrated in the convergence of optical spectra of organic solutes~\cite{Milanese2017} as well as VBEs~\cite{Cauet2010}.
By applying this approach to the benzene radical anion specifically, we aim at calculation of the VBE of its excess electron as a function of the cluster size ranging from a few solvent molecules to extensive clusters containing several thousand solvent molecules.
To be able to afford the calculations of the largest clusters, we mitigate the cost of the full DFT calculations by employing the methods of quantum mechanics and molecular mechanics (QM-MM)~\cite{Warshel1976, Hofer2018} and effective fragment potentials (QM-EFP)~\cite{Gordon2001, Gordon2012, Kaliman2013, Ghosh2013}.
Such dual interaction representation of the explicitly solvated system at the QM-MM level was shown to correspond well with the PCM alternative for the calculation of excitation energies~\cite{Provorse2016}. 
The results presented in this work shed light on the stability of the solvated benzene radical anion and thus provide a more rounded understanding of its solvent-induced stabilization.

\section{Computational Methodology\label{sec:methodology}}

\subsection{Cluster preparation}
\begin{figure}[tb]
    \centering
    \includegraphics[width=\linewidth]{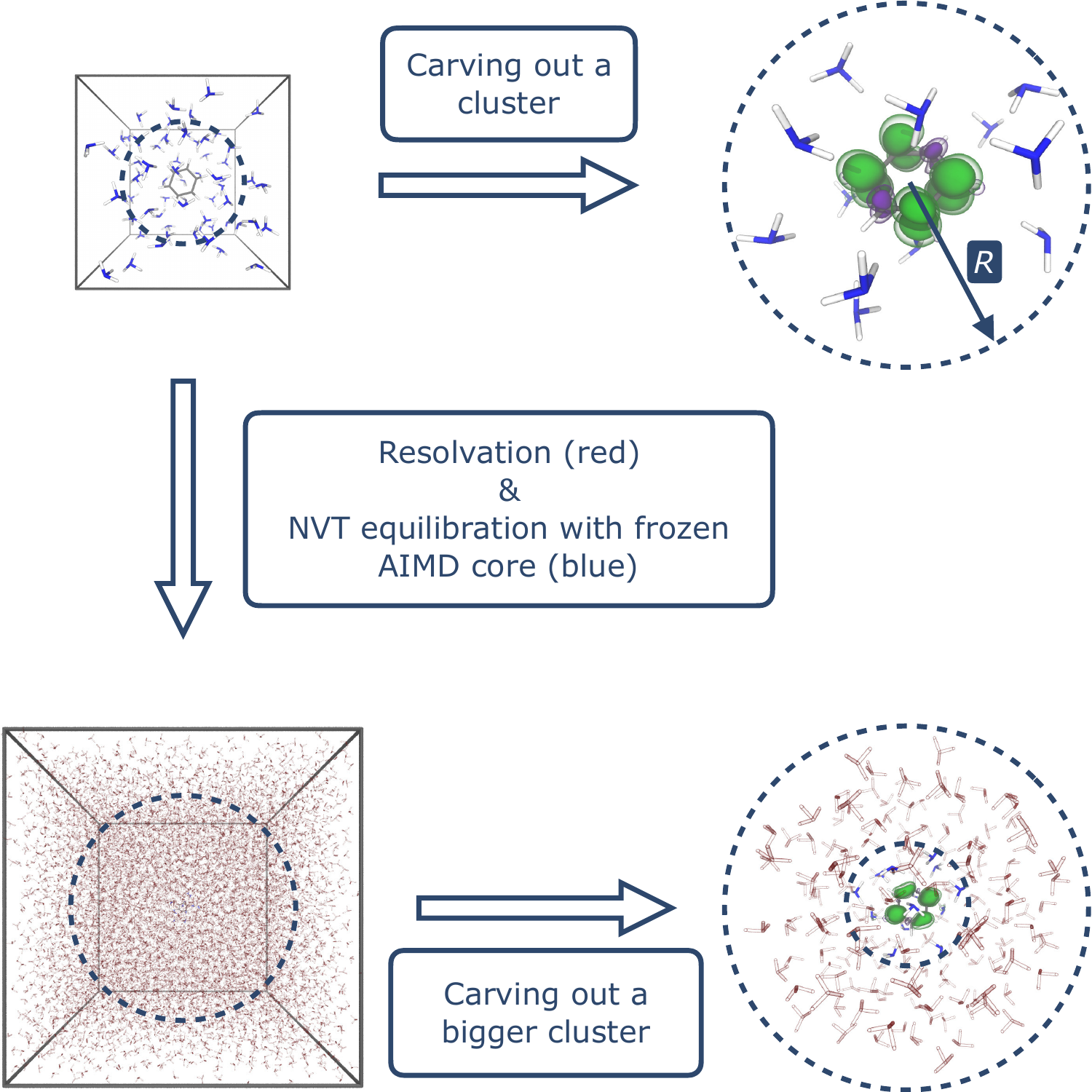}
    \caption{Schematic illustration of the cluster-carving procedure. Direct carving out of the AIMD simulation box with respect to the cutoff radius $R$ is shown in the top half of the figure.
    The resolvation and subsequent generation of a larger cluster is shown in the bottom half. 
    In the depicted clusters, the spin density contours are shown. 
    Green represents its positive part and violet the negative part at isovalues of 0.004~\AA$^{-3}$ (opaque) and 0.002~\AA$^{-3}$ (transparent) with respective signs.}
    \label{fig:Cluster_Scheme}
\end{figure}

The cluster structures were extracted from a bulk AIMD simulation in periodic boundary conditions~\cite{Brezina2020}.
Each cluster contained the benzene radical anion surrounded by explicit ammonia molecules up to a chosen cutoff radius of 6.8~\AA\ (Figure~\ref{fig:Cluster_Scheme}, top half); this was defined by the distance from the center of mass of the solute to the furthermost nitrogen atom in the solvation sphere.


To explore the properties of clusters larger than the size accessible by the original AIMD data, we employed a resolvation method in which each original structure was extended by additional solvent molecules and thermally equilibrated.
In particular, the core structure (defined by the above cutoff radius) was adopted from the original AIMD trajectory as before and centered in an empty extended cubic simulation box of side length of 73.85~\AA.
One then randomly placed additional solvent molecules into the unit cell around the core so that the total number of solvent molecules in the system was 10 000 while respecting the experimental density of liquid ammonia at 223~K~\cite{DonaldR.Burgess2018} combined with the previously estimated excluded volume of the benzene radical anion~\cite{Brezina2020}.
The new larger system was then equilibrated in the NVT ensemble using molecular dynamics with empirical force fields~\cite{Eckl2008, Wang200410.1002/jcc.20035}, however, with the original core kept constrained (additional information is provided in Section~S2 of the Supporting Information (SI)).
Following the equilibration, clusters of various cutoff radii were then carved out of the resolvated system consistently with the approach used for the original smaller systems (Figure~\ref{fig:Cluster_Scheme}, bottom half).
Alongside the thermal clusters sampled directly from the NVT trajectories, we prepared also optimized clusters for comparison. 
These included from zero to seven solvent molecules, with their initial configurations drawn from the original AIMD trajectory and their geometry minimized.

\subsection{Electronic Structure Calculations}
The VBE of the excess electron of the benzene radical anion in small clusters was calculated directly as the vertical difference between the total electronic ground-state energy of the anion and the neutral system at the same geometry in open boundary conditions. 
As such, it has a negative value for bound species; note that the often used term vertical detachment energy (VDE) is the same as VBE except for a sign change.
The electronic structure was characterized at the hybrid DFT level using the Q-Chem~5.3.2 software package~\cite{Shao2015}.
The revPBE0-D3~\cite{Perdew1996, Perdew1996a, Adamo1999, Grimme2011} functional together with the Ahlrichs type def2-TZVP basis set~\cite{Weigend2005} were employed.
This particular functional and basis set combination was chosen based on a methodological benchmark that is discussed in detail in the SI (Figure~S1).
Geometry optimizations were performed at the same hybrid DFT level of theory.
For the larger resolvated clusters, the cost of the full DFT calculations becomes very high, if not prohibitive. 
To overcome this problem, these systems were treated at a combined level of theory benefiting from the accurate quantum-mechanical (QM) description of the core that includes the computationally challenging radical anion and an affordable description of the distant solvent.
This was achieved in two different ways.
Within the first approach, the solvent outside the core was represented only by the partial point charges adopted from the rigid liquid ammonia force field introduced in Reference~\citenum{Eckl2008}.
In this QM-MM approach, the external solvent provides the key electrostatic effect on the QM core while not being electronically polarizable.
Alternatively, the more expensive combination of DFT description of the core with the effective fragment potential (EFP) to represent the distant solvent was employed to reach a similar goal.
Unlike QM-MM, the QM-EFP approach accounts for additional energy contributions in the outer solvent shell including exchange, polarization, and dispersion terms.

\subsection{Polarizable Continuum Models}
The gas-phase clusters of increasing size are used to model the progression from the isolated state towards the bulk.
PCM was used to represent the condensed phase in the context of cluster calculations to provide a reference bulk VBE for the series of clusters of increasing size as well as to explore the bulk properties in a computationally favorable way.
Two different versions of PCM are discussed in this work---the integral equation formalism (IEF-PCM)~\cite{Cances1997, Tomasi1999, Chipman2000} and the simplified conductor-like CPCM~\cite{Barone1998, Cossi2003}.
To account for the vertical electron detachment process, the non-equilibrium PCM was used, as implemented in Q-Chem~\cite{You2015}, to allow for the fast electronic polarization (but not for the slow nuclear polarization) of the PCM cavity after ionization.

In PCM calculations of non-covalently bonded systems, the construction of the cavity is of special importance since it can have a strong effect on the result.
Thus, three different possible approaches were compared.
The simplest molecular-shaped cavity formulation relies on overlapping atom-centered van der Waals (vdW) spheres scaled by an arbitrary factor, typically 1.2, which is used as default in Q-Chem.
Two more involved approaches are based on smoothing the vdW cavity with a probe sphere. 
The solvent excluded surface (SES) directly employs the original vdW surface while smoothing the sharp edges that arise due to the vdW sphere intersections. 
The solvent accessible surface (SAS) augments the cavity by adding the probe sphere radius.
Details about the cavity construction mechanisms were recently reviewed by Lange et al.~\cite{Lange2020}

\section{Results\label{sec:results}}

\subsection{Bulk Value of Vertical Binding Energy}

\begin{figure}[h]
    \centering
    \includegraphics[width=\linewidth]{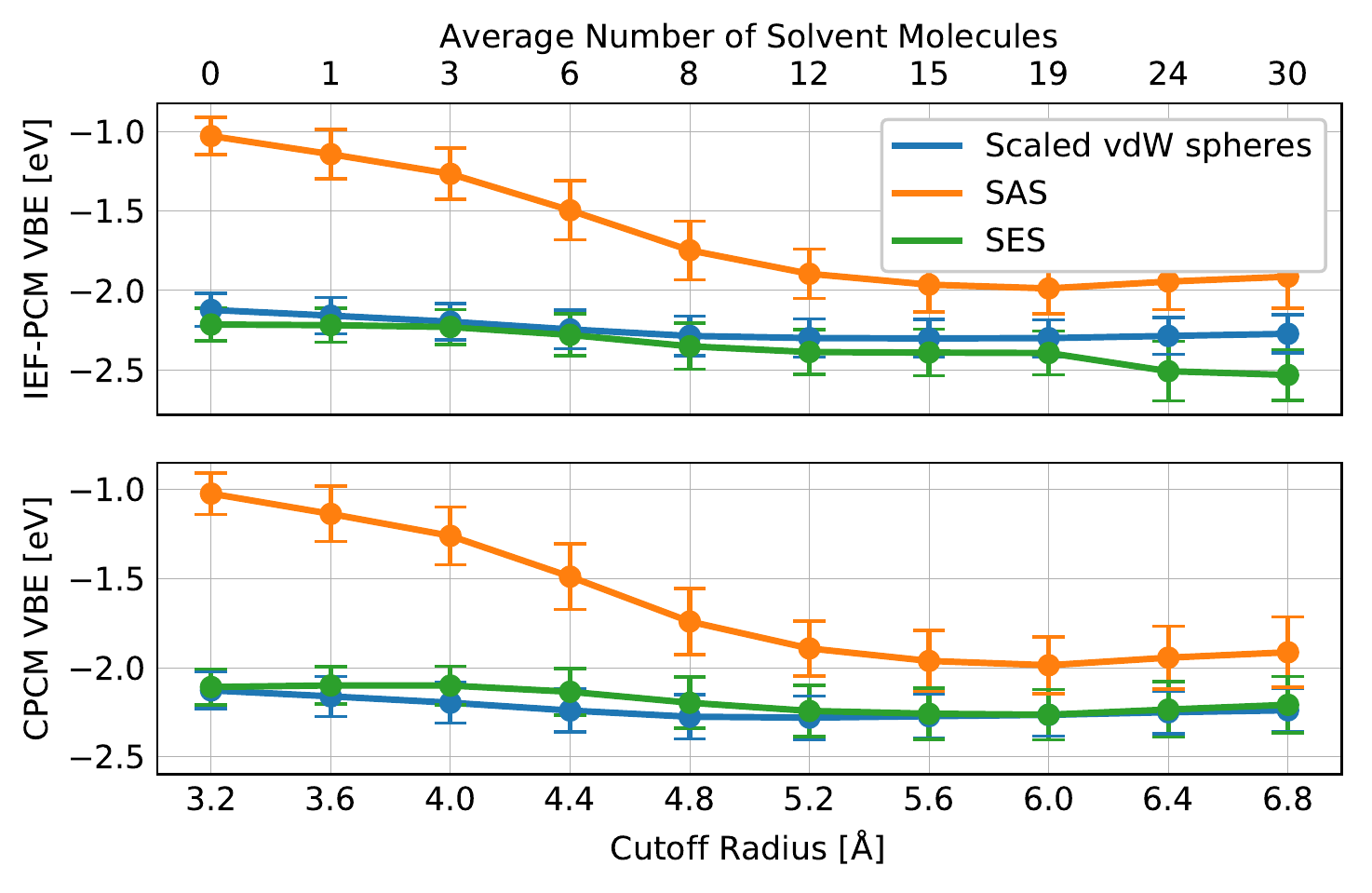}
    \caption
    {
    VBEs of the microsolvated clusters of different sizes embedded in the polarizable continuum and calculated by the IEF-PCM (top half) and by the CPCM (bottom half) methodology.
    Different cavity construction mechanisms are compared for each method.
    }
    \label{fig:PCM_BE}
\end{figure}

The VBEs in the limit of the bulk solvent environment were evaluated using PCM combined with explicit solvation up to the cutoff radius of 6.8~\AA\ with a set of 50 thermal configurations used for each data point.
This combination benefits from the efficient description of the bulk dielectric environment by the PCM and the inclusion of specific molecular interactions in close proximity to the benzene radical anion.

\begin{figure}[b]
    \centering
    \includegraphics[width=\linewidth]{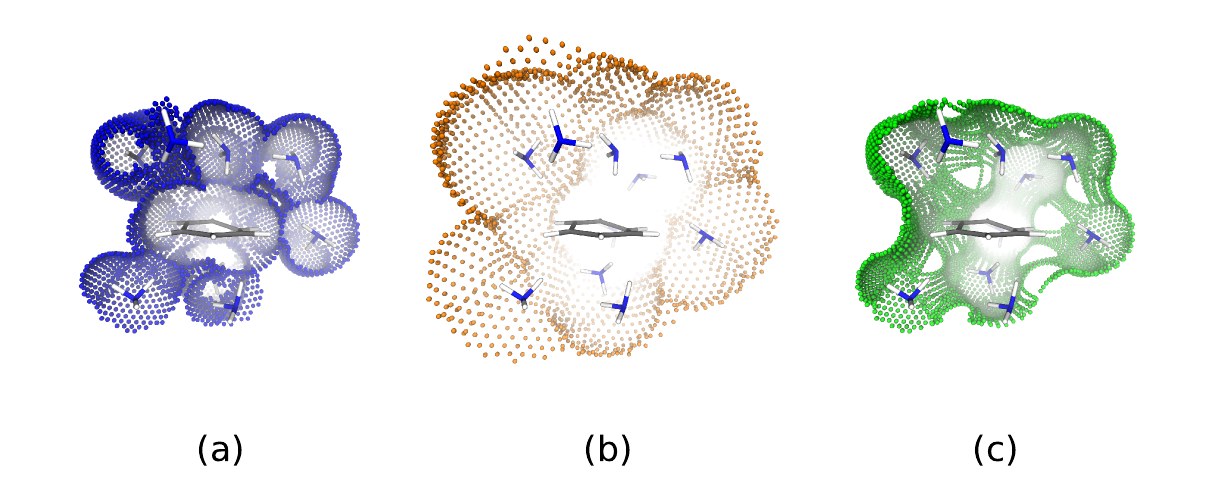}
    \caption
    {
    Visualization of the PCM cavities (colored surface points) constructed by different mechanisms at the same microsolvated cluster geometry:
    (a) Overlapping vdW spheres scaled by a factor of 1.2, (b) SAS, (c) SES.
    Each point of the cavity represents position of the partial charge at the cavity surface.
    }
    \label{fig:PCM_cavities}
\end{figure}

The VBE as a function of the size of the cluster of explicit ammonia molecules embedded in PCM is presented in Figure~\ref{fig:PCM_BE} for both the CPCM and IEF-PCM formulations, as well as for different types of the PCM cavities.
A representative example of the spatial distribution of the cavities is shown in Figure~\ref{fig:PCM_cavities}.
The resulting six chosen methods produce results consistent between the two PCM formulations but differing for the individual cavity construction mechanisms. 
The application of the most basic cavity construction mechanism relying on overlapping scaled vdW spheres leads to the VBE curves presented in blue in Figure~\ref{fig:PCM_BE}.
Here, we observe only a minor dependence of the PCM VBE on the cluster size with a mean value of $-$2.2~eV over the range of cutoff radii.
This corresponds to a situation in which the dielectric continuum alone already covers the major effects on the VBE and the additional explicit solvation adds only a minor correction.
However, a visual inspection of the shape of the vdW cavity (Figure~\ref{fig:PCM_cavities}a) indicates the presence of problematic regions where the PCM point charges penetrate between the molecules which has been previously reported~\cite{Simm2020} as a computational artifact that can lead to erroneous results.
The SAS cavity (orange curves in Figure~\ref{fig:PCM_BE}) is the most pronounced outlier that yields a VBE value as high as $-$1.0~eV for the isolated radical anion in PCM and then exhibits a noticeable, but slow drop to the value of roughly $-$2.0~eV for the longest cutoff radii.
These underbinding issues can be related to the excessive size of the SAS cavity which is formed by adding the probe sphere radius outline to an underlying vdW cavity.
This is illustrated by a snapshot of the cavity in Figure~\ref{fig:PCM_cavities}b. 
Clearly, this effect is the strongest for the isolated anion which represents the smallest system: as the cluster size grows by adding more solvent, the additional buffer layer between the edge of the molecular system and the cavity surface due to the added probe radius becomes less and less important.
The best of the two worlds is combined in the SES cavity (Figure~\ref{fig:PCM_cavities}c), which is wrapped tightly around the system like its vdW counterpart but also eliminates its unphysical attributes including the wrong size and the penetration inside the system.
As a result, in this case, we observe a VBE almost independent on the cluster size with its average value of $-$2.3~eV taken again over the calculated range of cutoff radii.

\subsection{Solvent-Induced Stabilization in Clusters}

\begin{figure}[h]
    \centering
    \includegraphics[width=\linewidth]{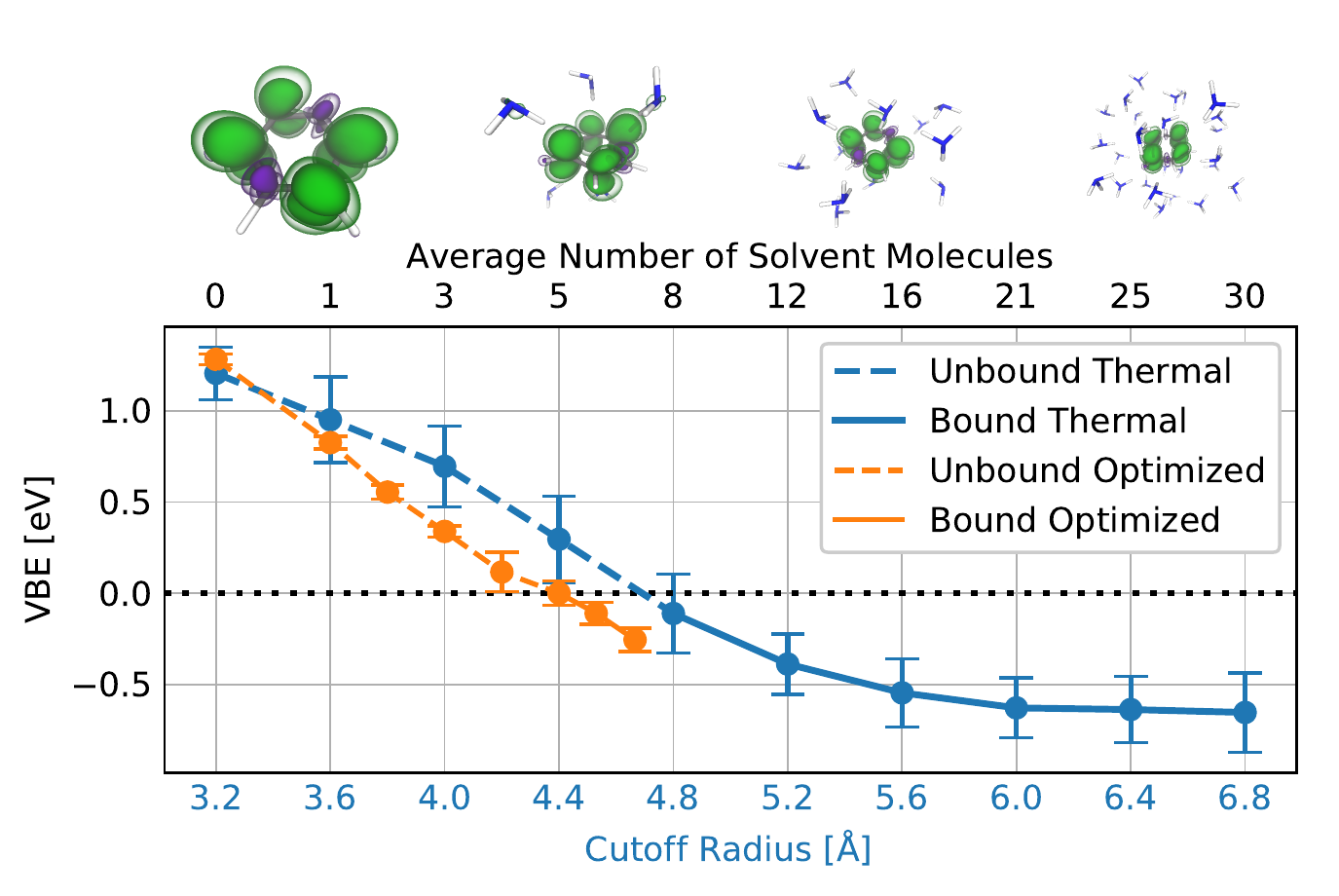}
    \caption
    {
    The dependence of the mean VBE on the cluster size together with the standard deviation.
    The dashed and full lines refer to the unbound (positive VBE values) and bound (negative VBE values) states, respectively.
    The VBEs of the thermal and optimized clusters are shown in blue and orange, respectively.
    Note that the bottom $x$-axis with cutoff radii relates to the thermal clusters only since the optimization changes the size and shape of the cluster.
    The top $x$-axis describes both the optimized and thermal structures.
    Typical examples of clusters with the corresponding spin densities at isovalues of 0.004~\AA$^{-3}$ (opaque) and 0.002~\AA$^{-3}$ (transparent) are presented at the top.
    }
    \label{fig:AIMD_clusters}
\end{figure}

Having set the PCM bulk baseline for the VBE values, we now address the question whether the same limit can be reached by gradually increasing the extent of solvation in isolated clusters.
The VBEs of clusters carved out from the AIMD trajectory were calculated for 50 thermal and 10 optimized structures at each cutoff radius and are depicted in Figure~\ref{fig:AIMD_clusters}.
The isolated benzene radical anion exhibits a positive value of VBE.
This means that the isolated species does not represent a bound electronic state but rather a resonance, consistent with the previous coupled cluster calculations~\cite{Bazante2015} and gas-phase experiments~\cite{Sanchef1973}. Strictly speaking, in such cases one should use methods pertinent to calculations of continuum states or resonances. 
When yielding positive values of VBE, the present calculations should thus be viewed as auxiliary only, since in the infinite basis set limit VBE would converge to zero.

A significant decrease of the VBE with the increasing cluster size can be seen from Figure~\ref{fig:AIMD_clusters}.
At a radius of 4.8~\AA\, corresponding on average to eight ammonia molecules, the mean VBE of thermal clusters crosses zero, indicating a transition to a stable electronic state.
A VBE value of approximately $-$0.6~eV is reached at a 6~\AA\ cutoff radius, followed by a near plateau in VBEs for larger clusters up to 6.8~\AA.
The emergence of this plateau coincides with the saturation of the first solvent shell at approximately 6.5~\AA\, as shown by the corresponding radial distribution functions (Figure~6 in Reference~\citenum{Brezina2020}).
An analogous trend is observed also for the optimized clusters with optimization leading to additional stabilization of the benzene radical anion. As a consequence, a bound state is reached already for 5 to 6 solvent molecules. (Figure~\ref{fig:AIMD_clusters}, orange curve).

\subsection{Resolvated Clusters}

\begin{figure}[b]
    \centering
    \includegraphics[width=\linewidth]{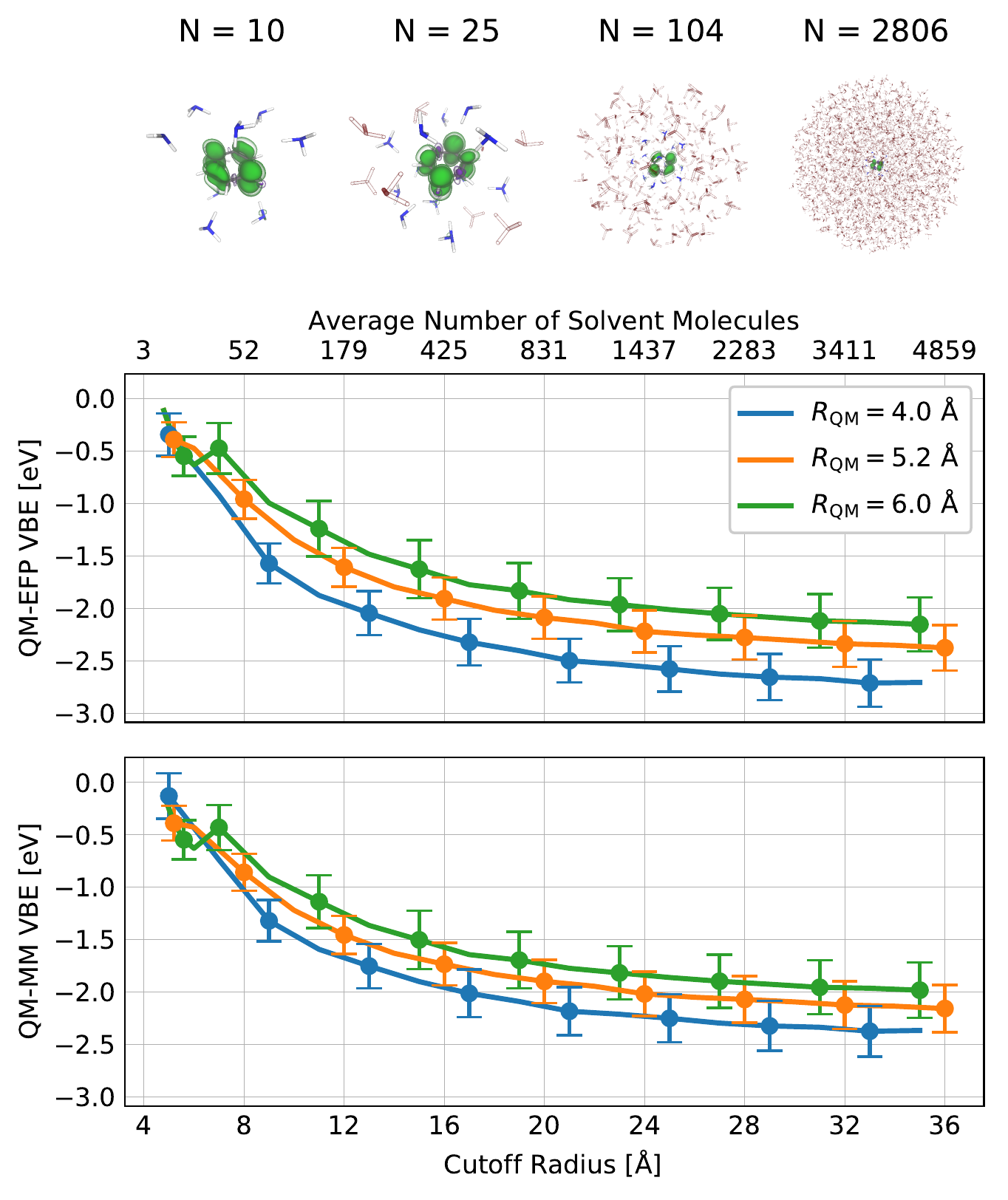}
    \caption
    {
    The VBEs calculated by QM-MM (bottom) or QM-EFP (middle) method for the resolvated clusters varying by the overall size (x-axis) and by the size of the inner QM subsystem (differently colored curves). 
    Typical geometries of resolvated clusters of increasing size are given in the top part where ammonia molecules belonging to the QM subsystem are depicted in white-blue and the rest in red.
    Additionally, the spin density is presented as green contours at isovalues of 0.004~\AA$^{-3}$ (opaque) and 0.002~\AA$^{-3}$ (transparent).
    }
    \label{fig:QM-MM-EFP}
\end{figure}

To generate larger clusters, our method of resolvation was used and the VBEs were calculated employing the two different approaches to system extension---QM-MM and QM-EFP---as described in Section~\ref{sec:methodology}.
The resulting dependencies of the VBE on both the size of the cluster as a whole and the size of the QM core subsystem are shown in Figure~\ref{fig:QM-MM-EFP}.
The QM-EFP approach smoothly converges with the cluster size towards VBE ranging from $-$2.0 to $-$2.6~eV for the largest cluster sizes depending slightly on the chosen size of the QM core.
The computationally simpler QM-MM approach displays a qualitatively similar trend and features a smaller dependence on the size of the QM core.
Here, the VBEs for the largest clusters range between $-$1.9 and $-$2.3~eV.
However, the good agreement between the two methods is likely fortuitous.
As documented in Section~S4 of the SI and discussed in detail in Section~\ref{sec:discussion}, we tested changing the parametrization by replacing the rigid ammonia model~\cite{Eckl2008} with a closely related flexible one~\cite{Engin2011}.
This features a slight modification of the partial point charges and, by definition, produces a distribution of different ammonia geometries.
Here, the quantitative VBE values as well as the asymptotic behavior change significantly for QM-MM, whereas the QM-EFP results remain largely unaffected.
Note also that both the QM-MM and QM-EFP curves display a small dent at a distance corresponding to the boundary of the QM core, which can be attributed to the seam between the subsystems described by different methods.

\begin{figure}[tb]
    \centering
    \includegraphics[width=\linewidth]{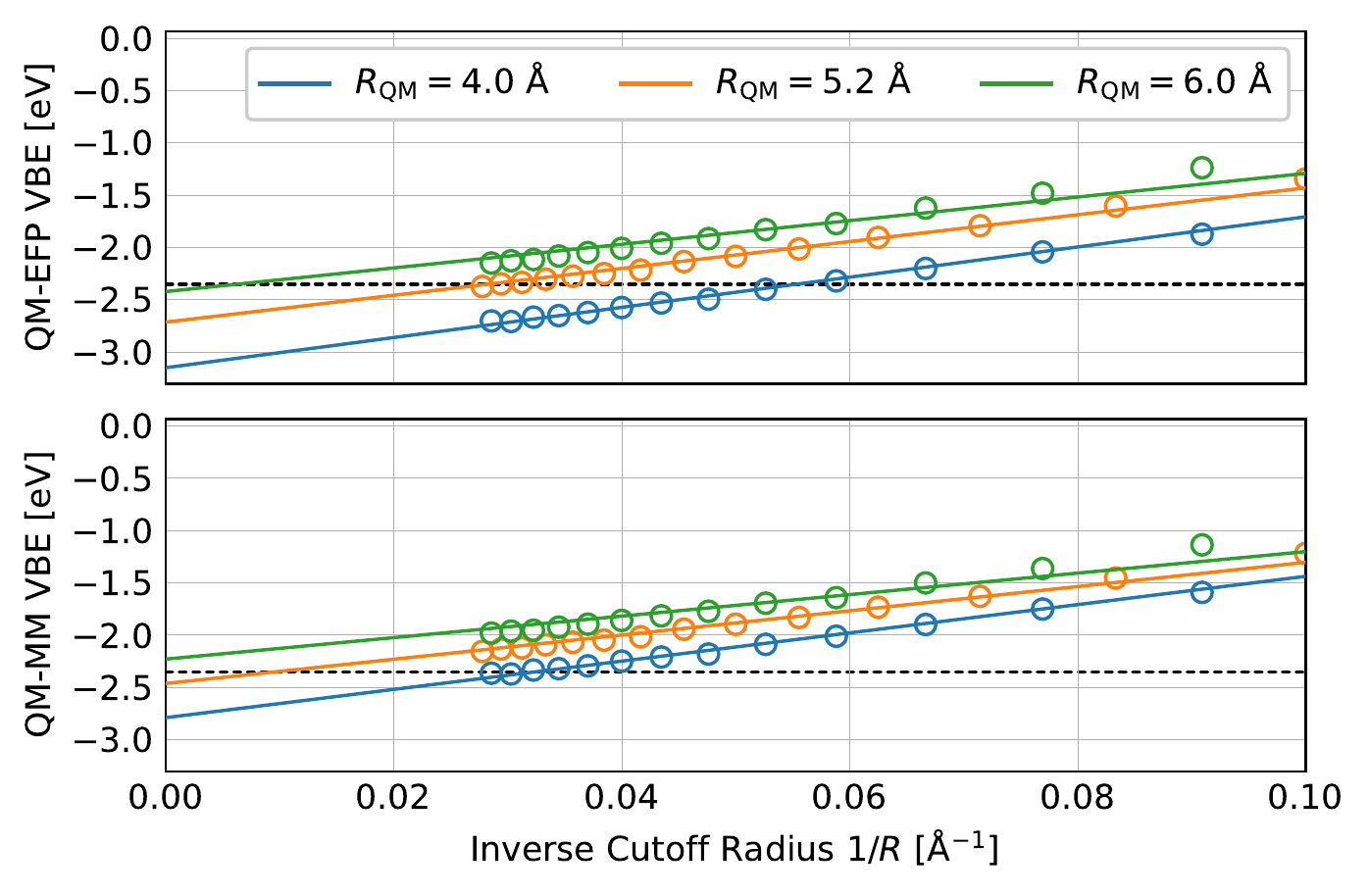}
    \caption
    {
    A plot of the VBEs obtained at the QM-EFP (top panel) and the QM-MM (bottom panel) level as a linear function of the reciprocal cutoff radius on the $x$-axis.
    The original data are shown in empty circles and the fitted curves as solid lines.
    The colors distinguish the particular QM subsystem sizes.
    The reference VBE value obtained from the IEF-PCM/SES calculation is shown as a black dashed line at $-2.35$~eV.
    }
    \label{fig:Resolv-VBE-fit}
\end{figure}

To gain insight into the limit of infinite cluster size, we fitted the VBEs with reciprocal functions.
While the specific parameters of the fit are presented in the Table~S1 of the SI, we show here the linear dependence on the reciprocal radius extrapolating the VBE to the large cluster limit in Figure~\ref{fig:Resolv-VBE-fit} where the limiting VBE values can be read off at the vertical axis intercept.
Here, the extrapolation introduces an additional minor decrease in the VBEs which now range between $-$2.3 and $-$3.1~eV for QM-EFP and between $-$2.1 and $-$2.7~eV for QM-MM.

\section{Discussion\label{sec:discussion}}


The use of clusters that include the benzene radical anion together with a small number of explicit ammonia molecules embedded in a polarizable continuum gives access to the VBE of the anion's excess electron in bulk solution while eliminating the need for difficult calculations in extended periodic systems.
Within the IEF-PCM/SES approach, the VBE amounts to $-2.3$~eV which implies that the benzene radical anion is relatively strongly bound relative to the vacuum level, being stabilized by the solvent. Notably, this VBE value is roughly reached already without any explicit ammonia molecules, which suggests that the continuum dielectric environment is sufficient to semi-quantitatively describe the solvent stabilization of the benzene radical anion.
This is also in line with the observed stability in other polar solvents that were used in experimental studies~\cite{Mortensen1984, Marasas2003} and that have similar dielectric constants to that of liquid ammonia.
The above result is in a quantitative agreement with more accurate but also more computationally demanding G$_0$W$_0$ calculations~\cite{Brezina2021}. Experimentally, this VBE value should be accessible using X-ray photoelectron spectroscopy (XPS) in refrigerated liquid microjets~\cite{Buttersack2020a}. 
Note that in previous studies, we have successfully exploited the combination of accurate G$_0$W$_0$ calculations with XPS to address the electronic structure of liquid ammonia~\cite{Buttersack2019} and electrons solvated in it~\cite{Buttersack2020}. 

Calculations of the benzene radical anion in ammonia clusters of increasing size complement the above PCM calculations by providing insight into the process of stabilization of the anion as the extent of solvation is increased.
The stabilization curves of VBE with respect to the cluster cutoff radius can be characterized by two points that describe the stabilization progress.
First is the size of the system where the binding energy transitions into negative values, meaning that the system becomes vertically bound.
For the benzene radical anion, the turning point occurs for clusters with 7-8 ammonia molecules.
This is consistent with previous observations of other species that are unstable in the gas phase, including the sulfate dianion~\cite{Wang2009} and the hydrated electron~\cite{CoeA1990, Marsalek2010a, Jacobson2011} that are both stabilized by several water molecules.
Interestingly, only a small difference (within the spread over individual clusters) in this turning point is observed between the above thermal clusters carved from AIMD simulations at 223~K and optimized clusters, the latter reaching negative VDEs at 5 to 6 ammonia molecules.
This implies that for the benzene radical anion in ammonia clusters the exact conditions of cluster preparation may have only a weak effect on the measured VBEs. 

The second important point concerns the asymptotic behavior in the large cluster limit.
In this work, this limit is approached by invoking a dual interaction representation where the outer solvent layer, which can contain a large number of molecules, is treated by a computationally less demanding method than the hybrid DFT used for the inner part of the cluster.
Notably, the QM-EFP method produces robust asymptotic VBE convergence to values between $-$2.3 and $-$3.1~eV, which correlates quantitatively with our bulk PCM value as well as our previous G$_0$W$_0$ calculations~\cite{Brezina2021}.
An extrapolated VBE value of exactly $-$2.3~eV, which is absolutely consistent with the other methods, is achieved with the QM-EFP method applied to the system with the largest QM core.
This suggests that a more accurate description of the VBE of the excess electron of the benzene radical anion is obtained when the solute is separated from the resolvation region by a full layer of QM solute which is not the case with the smaller QM core sizes.
As a general trend, our results obtained from both the QM-EFP and QM-MM methods point to the fact that the convergence with cluster size is very slow, which is due to the long-range character of the solvent polarization by the central ion~\cite{Cauet2010, Jagoda-Cwiklik2008}. 
The negative slope of the VBE curves starts to level off only at the cutoff radius of 36~\AA\ where the clusters already contain several thousands of solvent molecules.


From the technical point of view, the results reported in this work are based on a set of computational methods that can be compared and cross-validated among each other as well as with other reference methods.
PCM is a favorite choice for including solvent effects into quantum-chemical calculations in a computationally inexpensive manner.
In this work, we explore the combinations of two state-of-the-art PCM formulations with three different PCM cavity construction mechanisms for VBE calculations of a solvated radical anion.
We find that the results depend on the particular choice of cavity construction and, therefore, must be subject to validation before drawing quantitative conclusions.
In our case, the SES cavity with either of the employed formulations yields results consistent both internally with the large size limit of the cluster series and with G$_0$W$_0$ calculations~\cite{Brezina2021}.
At the same time, the SES approach was previously identified as a reliable formulation of the PCM cavity for the purpose of calculations of excitation energies in the presence of explicit solvent molecules~\cite{Provorse2016}.
Surprisingly, we find that the simplest vdW cavity yields results comparable with SES, however, this may be to some extent due to error compensation.
The fact that the present results are in agreement with G$_0$W$_0$ calculations~\cite{Brezina2021} serves as useful cross-validation. 
G$_0$W$_0$ performed in periodic systems does not directly provide absolute VBE values and requires additional alignment, in this case using the valence band of the solvent, so these PCM calculations provide additional verification.
At the same time, as a highly accurate method that does not depend substantially on the density functional that provides the underlying orbitals~\cite{Wilhelm2016}, G$_0$W$_0$ justifies the choice of the density functional used in the present VBE calculations.

The dual interaction scheme used for calculations of extended clusters is realized here in two different ways, as a QM-MM or a QM-EFP scheme.
Both appear to produce VBE values consistent with each other as well as with the principally different computational methods based on G$_0$W$_0$, but the QM-EFP approach seems to be more robust and generalizable.
The specific manifestation of this is represented by the fact that the QM-MM results seem consistent with QM-EFP only with a particular choice of a model, but change significantly once the parametrization of the solvent is modified.
This showcases the high sensitivity of the QM-MM method to the chosen force field and thus favors the QM-EFP results in which the fixed fragment geometry and its parametrization buffer a large part of the changes that are experienced when only bare point charges are used.
Note that the computational expense of calculating the systems is given largely by the size of the QM core itself; the additional computational overhead is negligible in the QM-MM case and modest for QM-EFP (Figure~S2). 
Therefore, the dual interaction representation, particularly its QM-EFP flavor, is found to be a practical and reasonably accurate choice for VBE calculations of very large clusters.

\section{Conclusion\label{sec:conclusion}}

The present work reports on investigations of the solvation-induced electronic stabilization of the benzene radical anion by quantum-chemical calculations of VBEs in ammonia clusters of increasing sizes and in the liquid bulk.
To this end, we employ hybrid DFT electronic structure calculations necessary to accurately describe the radical anion, augmented by either the QM-MM or the QM-EFP scheme in order to describe large clusters.
The reported calculations represent a computationally viable way to calculate the VBE of the excess electron of the benzene radical anion in liquid ammonia.
First, we show by calculations of small clusters carved from an AIMD trajectory and embedded in PCM that the benzene radical anion is electronically stabilized by solvation in bulk ammonia with the VBE value reaching $-$2.3~eV.
Second, progression toward this value is demonstrated for a series of clusters of increasing size.
Here, we find that a a small number of 5 to 8 ammonia molecules is sufficient to vertically stabilize the benzene radical anion, while several thousand molecules are needed to converge to the above bulk value of VBE.
The present study thus provides a detailed view of the process of electronic stabilization of the benzene radical anion in ammonia upon extending the solvation environment.
It also provides VBE estimates experimentally verifiable via photoelectron spectroscopy with a high degree of consistency between the employed computational methods.

\section*{Supporting Information}

Additional computational details and benchmarks.
Detailed description of the resolvation technique.
Extrapolation parameters of the VBE dependencies in resolvated clusters to the large cluster limit.
Results and discussion of VBE differences induced by employing a flexible force field for QM-MM and QM-EFP calculations.

\section*{Acknowledgment}

P.J. acknowledges support from the European Regional Development Fund (Project ChemBioDrug no. CZ.02.1.01/0.0/0.0/16\_019/0000729) and the Humboldt Research Award.
K.B. acknowledges funding from the IMPRS for Many Particle Systems in Structured Environments.
This work was supported by the Project SVV 260586 of Charles University.
This work was partially supported by the OP RDE project (No. CZ.02.2.69/0.0/0.0/18\_070/0010462), International mobility of researchers at Charles University (MSCA- IF II).

\end{document}


\def\mytitle{Supporting Information for: Benzene Radical Anion Microsolvated in Ammonia Clusters: Modelling the Transition from an Unbound Resonance to a Bound Species}
\title{\mytitle}

\author{Vojtech Kostal}
\affiliation{
Institute of Organic Chemistry and Biochemistry of the Czech Academy of Sciences, Flemingovo nám. 2, 166 10 Prague 6, Czech Republic
}

\author{Krystof Brezina}
\affiliation{
Charles University, Faculty of Mathematics and Physics, Ke Karlovu 3, 121 16 Prague 2, Czech Republic
}
\affiliation{
Institute of Organic Chemistry and Biochemistry of the Czech Academy of Sciences, Flemingovo nám. 2, 166 10 Prague 6, Czech Republic
}

\author{Ondrej Marsalek}
\affiliation{
Charles University, Faculty of Mathematics and Physics, Ke Karlovu 3, 121 16 Prague 2, Czech Republic
}

\author{Pavel Jungwirth*}
\email{pavel.jungwirth@uochb.cas.cz}
\affiliation{
Institute of Organic Chemistry and Biochemistry of the Czech Academy of Sciences, Flemingovo nám. 2, 166 10 Prague 6, Czech Republic
}

\date{\today}

\maketitle

\section{Basis set benchmark}

\begin{figure}[tb]
    \centering
    \includegraphics[width=\linewidth]{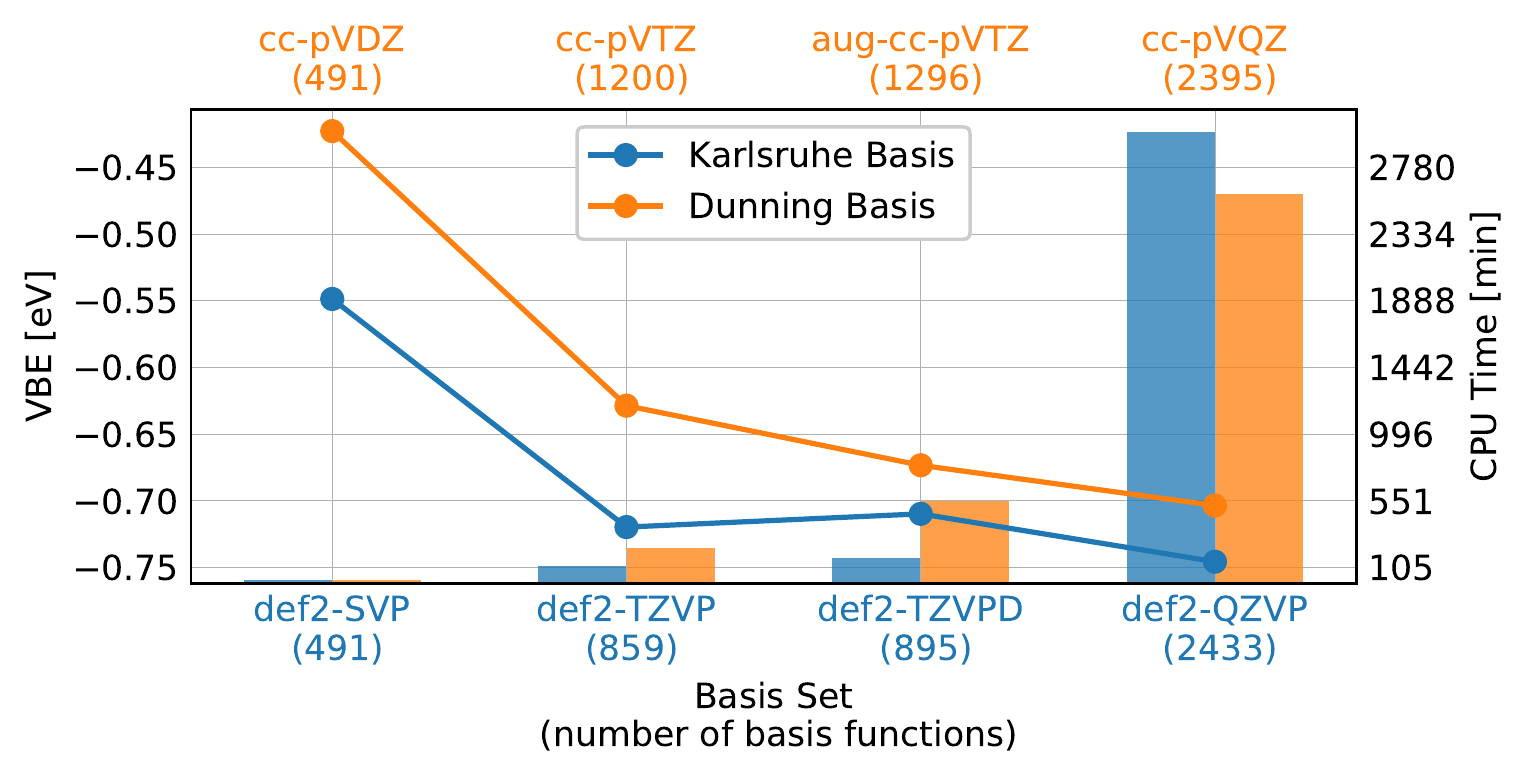}
    \caption
    {
    A benchmark of Karlsruhe (blue) and Dunning (orange) basis sets performance in combination  with the revPBE0-D3 density functional.
    The binding energies are plotted as curves and their values are shown at the left-hand side $y$-axis.
    The corresponding CPU times are depicted as bars with the same color coding.
    Moreover, for each basis set employed, the number of basis set functions is given in brackets under the basis set names on the top and bottom $x$-axes.
    Note that the augmentation with diffuse functions (def2-TZVPD and aug-cc-pVTZ) is applied only on carbon atoms.
    }
    \label{fig:bs_benchmark}
\end{figure}

The disperssion-corrected revPBE0-D3~\cite{Perdew1996, Perdew1996a, Zhang1998, Adamo1999, Goerigk2011, Grimme2010} density functional was used in line with our previous work on the benzene radical anion where we have documented the need for a hybrid functional to overcome the unphysical delocalization of the excess electron that is observed at the generalized-gradient-approximation level~\cite{Brezina2020}. 
Using this functional, the effect of the basis set size on the calculated VBEs was benchmarked.
The VBEs were calculated for a randomly chosen cluster consisting of the benzene radical anion surrounded by 13 ammonia molecules.
At this size, the system is already electronically bound and the corresponding VBE values are thus quantitative.
As shown in Figure~\ref{fig:bs_benchmark}, Dunning type~\cite{Dunning1989} basis sets provided roughly similar results to those obtained using Karlshure basis sets~\cite{Weigend2005}.
The use of triple-$\zeta$ basis sets lead to a significant improvement in comparison to the smaller double-$\zeta$ ones.
A further augmentation by diffuse (def2-TZVPD and aug-cc-pVTZ) or polarization functions (def2-QZVP and cc-pVQZ) resulted in an non-significant VBE decrease of less than 0.1~eV but in an order-of-magnitude elevation of the computational cost.
Therefore, we conclude that the def2-TZVP basis set represents the best compromise between the accuracy of the method and its computational requirements.

\section{Resolvation details}

\subsection*{Force-Field Molecular Dynamics Details}

All auxiliary MD simulations were realized in the Gromacs~2020.4~\cite{Berendsen1995, Lindahl2020} software.
A 100~ps long NVT equilibration run was performed at 223~K employing a 0.5~fs integration time step and the stochastic velocity rescaling thermostat~\cite{Bussi2007}.
Energy and forces were evaluated using the rigid force field for liquid ammonia~\cite{Eckl2008} and the generalized Amber force field describing the benzene radical anion~\cite{Wang200410.1002/jcc.20035}.

The method of resolvation relies on several steps that are summarized below.
\begin{enumerate}
    \item Carve a spherical cluster out of the original periodic AIMD trajectory frame.
    Its size is defined by distance between the center of mass of the benzene radical anion and the furthermost ammonia nitrogen atom within the cluster.
    \item Determine the Mulliken partial charges~\cite{Mulliken1955} at all atoms of a specific benzene structure at the revPBE0-D3/def2-TZVP level.
    \item Modify the topology of benzene radical anion with the obtained Mulliken charges.
    \item Center the cluster from the point~1 into a cubic unit cell with 73.85~\AA\ side length.
    \item Fill this box with ammonia molecules randomly placed around the core cluster such that the minimal distance between atoms is not less than 2~\AA.
    \item Equilibrate the system using NVT molecular dynamics with the original core constrained.
    \item Carve a resolvated cluster out of the last NVT equilibration trajectory frame with respect to an overall cutoff radius defining the cluster size including the inner core.
\end{enumerate}

\subsection*{Resolvation CPU-time scaling}

\begin{figure}[tb]
    \centering
    \includegraphics[width=\linewidth]{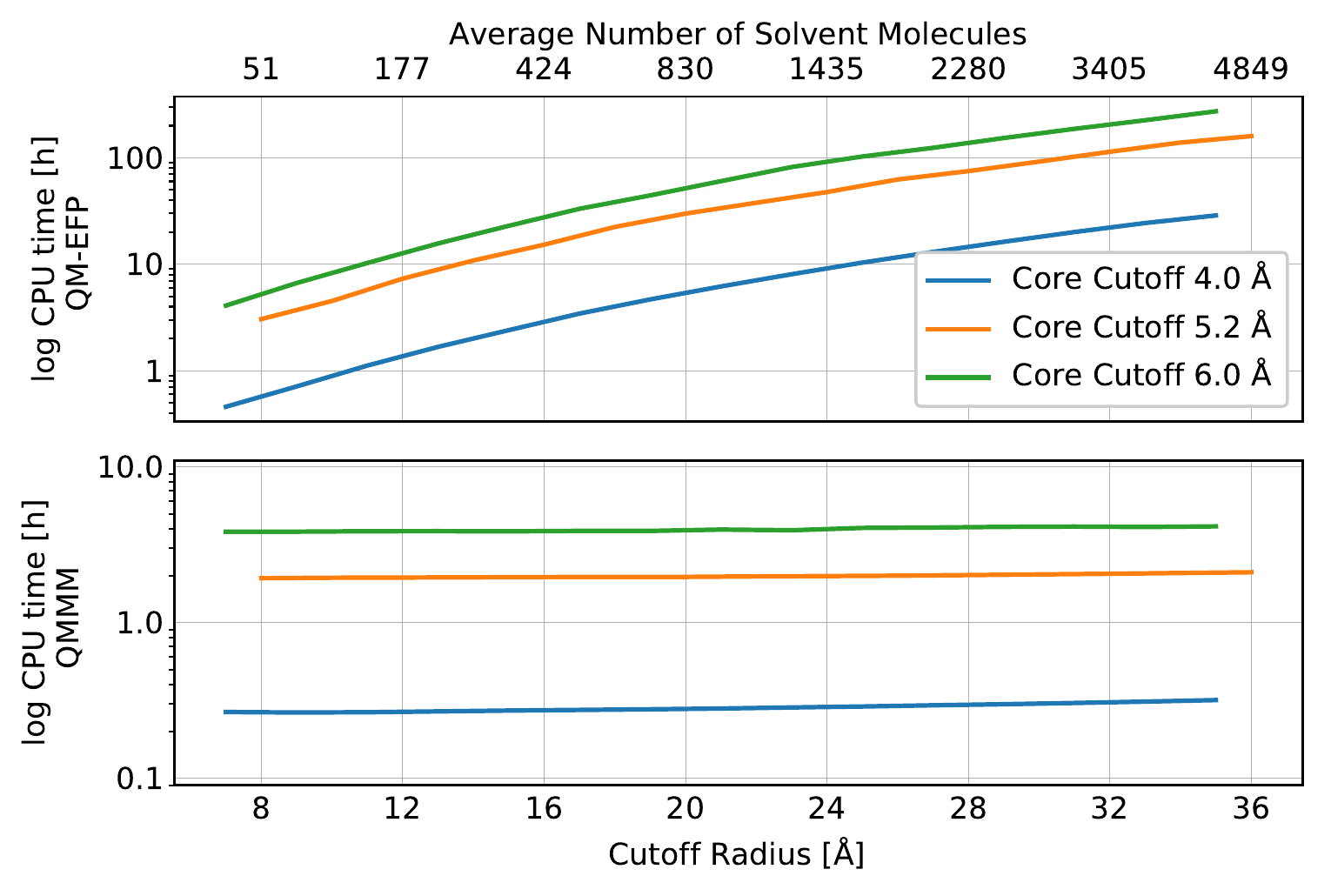}
    \caption{The dependence of the CPU time needed to perform either a QM-EFP (top) or a QM-MM (bottom) calculation of the VBE on the cluster size.
    Different QM subsystem sizes are distinguished by the color scheme used consistently in the main text and explained in the legend.}
    \label{fig:QM-MM-EFP-CPUtime}
\end{figure}

Scaling of the computational cost for methods of QM-MM and QM-EFP is shown in Figure~\ref{fig:QM-MM-EFP-CPUtime} as a function of the overall cutoff radius of the resolvated cluster.

\section{Parameters of the Extrapolated VBE curves}

\begin{table}[h]
    \centering
    \begin{tabular}{ccccc}
        \toprule
        \multirow{2}{*}{Core Cutoff [\AA]}
        & \multicolumn{2}{c}{QM-EFP} & \multicolumn{2}{c}{QM-MM} \\
        & $k$ & $q$ & $k$ & $q$ \\
        \midrule
        4.0 & 14.437 & $-$3.151 & 13.516 & $-$2.790 \\
        5.2 & 12.837 & $-$2.715 & 11.590 & $-$2.463 \\
        6.0 & 11.306 & $-$2.423 & 10.283 & $-$2.230\\
        \bottomrule
    \end{tabular}
    \caption{Parameters $k$ and $q$ of the fitted solid lines in Figure~6 given as $k/R+q$.
    Note that the parameter $q$ corresponds to the VBE limit when $R\rightarrow\infty$.}
    \label{tab:fit-params}
\end{table}

\section{Resolvation Employing the flexible Ammonia Force Field}

\begin{figure}[h]
    \centering
    \includegraphics[width=\linewidth]{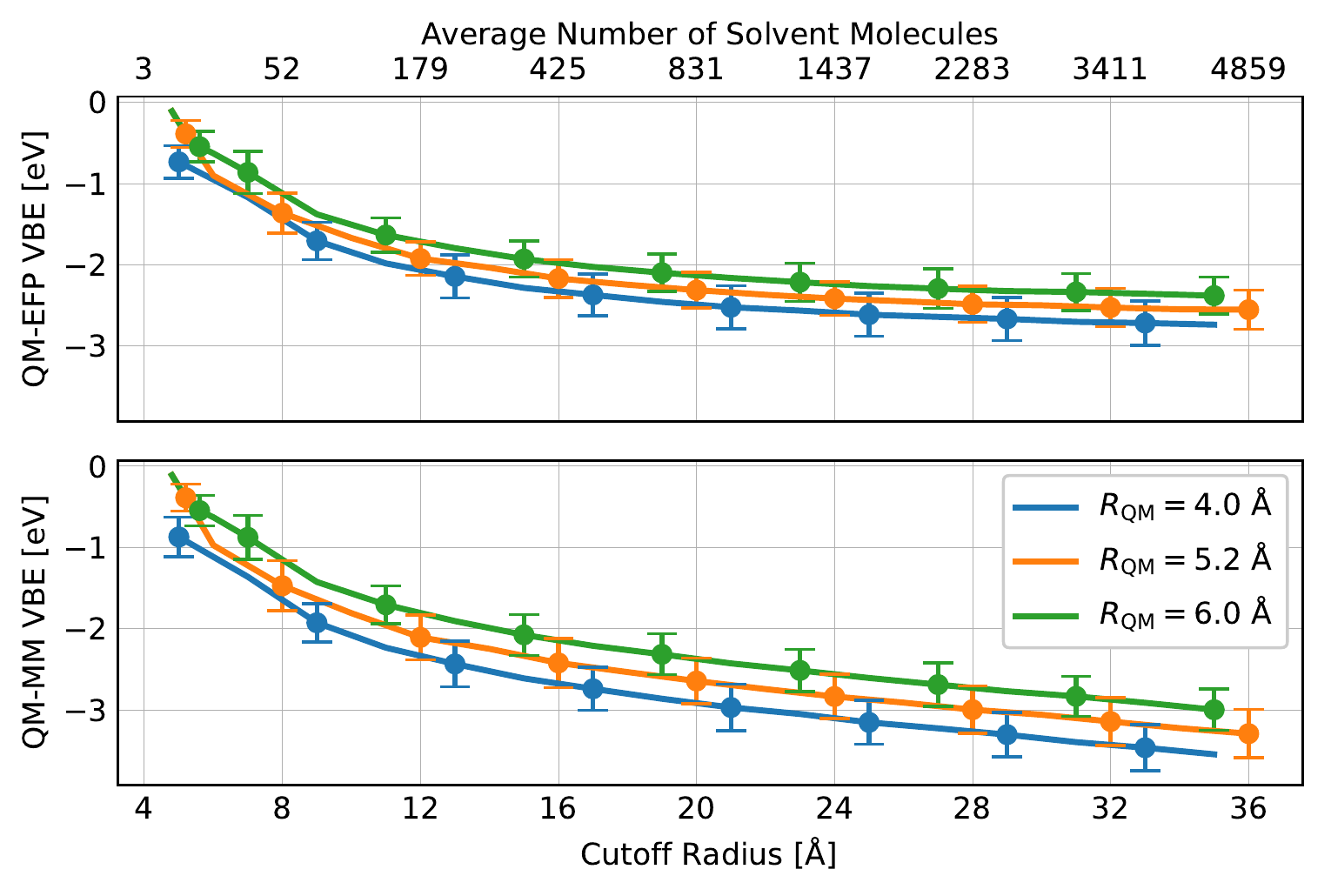}
    \caption
    {
    The mean VBEs of the clusters resolvated using the flexible ammonia force field~\cite{Engin2011} as a function of the overall cutoff radius (bottom $x$-axis) as well as the inner core cutoff which is defined by the color coding.
    The QM-EFP curves are shown in the top panel and the QM-MM curves in the bottom panel.
    Representative numbers of average number of ammonia molecules in the clusters are given in the upper $x$-axis.
    }
    \label{fig:vbe-qm-efp-mm-flexible}
\end{figure}

The VBE curves obtained from resolvation by the flexible liquid ammonia force field~\cite{Engin2011} are presented in Figure~\ref{fig:vbe-qm-efp-mm-flexible}.
Note the pronounced difference between the QM-MM curves here and those presented in Figure~5 of the main text.

\section{PCM details}

The non-equlibrium PCM was based on the Marcus partition scheme~\cite{You2015}.
All PCM cavities were discretized into 194 surface points for the heavy atoms while for hydrogens we used 194 points per atom in the SES case and 110 for the vdW and SAS surfaces.
The liquid ammonia  environment was characterized for the non-equilibrium PCM purposes by a pair of dielectric constants splitting the solvent response into a slow (nuclear) and fast (electronic) parts: the low-frequency one of 22.66 and the high-frequency one of 1.9444~\cite{JohnR.Rumble2020}.

\renewcommand{\bibsection}{%